# Direct imaging of plasmonic coupling between gold nanoparticles and one-dimensional van der Waals materials

Ibrahim Tanriover[1,#], Arindom Nag[2,#], Jong-Ho Lee[2], Huy Hoang[2], Haihua Liu[1,*] and Thang Pham[2,*]

[1]Electron and X-ray Microscopy Group, Center for Nanoscale Materials, Argonne National Laboratory, Lemont, IL 60439, USA
[2]Department of Materials Science and Engineering, Virginia Polytechnic Institute and State University, Blacksburg, VA 24060, USA
[#]: These authors contributed equally to this work
*Corresponding Authors: haihua.liu@anl.gov, thangpham@vt.edu

Abstract:

Hybrids of plasmonic metals and van der Waals (vdW) materials combine electromagnetic field enhancement with the electronic properties of the vdW component, but the studies reported so far concern two-dimensional layers, and the near field at the metal/vdW interface is generally inferred from far-field measurements. Here we report the plasmonic coupling between Au nanoparticles and a one-dimensional (1D) vdW material, $NbS_3$, and measure the near field at the Au-$NbS_3$ contact directly. $NbS_3$ nanowires and nanoribbons were grown by salt-assisted vapor-liquid-solid growth on graphene-covered TEM grids, and atomic-resolution electron microscopy shows atomically sharp Au/$NbS_3$ interfaces. Photon-induced near-field electron microscopy at 1030 nm shows that an Au particle in contact with $NbS_3$ produces a near field of higher intensity and larger extent than an isolated particle of similar size, and that the hot spot at a contact is excited only when the incident polarization is parallel to the particle-$NbS_3$ axis. Finite-difference time-domain simulations reproduce these observations and quantify the dependence of the coupling on the particle-$NbS_3$ gap, the $NbS_3$ thickness, and the dimension ratio of the $NbS_3$ structure. Size-specific Au nanoparticles on $NbS_3$ nanowires enhance the Raman scattering of rhodamine 6G by a factor of 3.2 ± 1.4 relative to Au nanoparticles alone, with a detection limit of $10^{-7}$ M. These results extend plasmonic metal/vdW hybrids to 1D vdW crystals and establish a multimodal approach that correlates the atomic structure of a metal/vdW interface with its nanoscale plasmonic response.

## 1. Introduction

Transition metal trichalcogenides (TMTs), $MX_3$ (M = Ti, Zr, Hf, Nb, Ta; X = S, Se, Te), are quasi-one-dimensional (1D) van der Waals (vdW) materials in which $MX_6$ trigonal prisms share faces to form infinite chains along the [010] axis[1–3]. Bonding is covalent within a chain and vdW between chains. This structural anisotropy results in direction-dependent electrical and optical properties[2,4]. Among TMTs, $NbS_3$ crystallizes in several polymorphs[5,6] whose electronic structure ranges from semiconducting (phases I and IV) to metallic (phases II, III, and V). The metallic phase II has been synthesized as nanowires and nanoribbons by salt-assisted vapor-liquid-solid (VLS) growth with Au-based catalysts and by catalyst-free vapor–solid (VS) growth[7,8]. The optical response of these nanostructures, and in particular their interaction with plasmonic metals, has not been investigated.

Localized surface plasmon resonances (LSPRs) of noble metal nanoparticles confine optical fields to dimensions well below the wavelength of light[9,10]. The resulting near-field enhancement is the basis of surface-enhanced Raman scattering (SERS), in which the electromagnetic enhancement factor scales approximately with the fourth power of the local field and reaches $10^4$-$10^8$ for noble metal substrates[11]. VdW materials can also enhance Raman scattering, mainly by charge transfer to the adsorbed molecule, but the enhancement factors are typically limited to $10^2$-$10^3$ [11]. Combining a plasmonic metal with a vdW material therefore provides both electromagnetic and chemical enhancement[11–13]. Reported examples include Au nanoparticles on $MoWS_2$, with a detection limit of $10^{-11}$ M for rhodamine 6G (R6G)[11], and 1T'-$WS_2$ monolayers on 4H-Au nanowires, with a detection limit of $10^{-18}$ M[14]. In these hybrids, the enhancement is determined by the near field at the metal/vdW interface. This near field is generally inferred from far-field Raman spectra and from electromagnetic simulations rather than measured[15,16]. Furthermore, the vdW components investigated to date are two-dimensional (2D) layers.

Photon-induced near-field electron microscopy (PINEM), a mode of ultrafast electron microscopy (UEM), provides a direct measurement of optically excited near fields in real space[17,18]. A femtosecond laser pulse excites the specimen, and a synchronized electron pulse traversing the evanescent field gains or loses integer multiples of the photon energy. An image formed from the electrons that gained energy maps the field component along the electron trajectory with nanometer spatial resolution and sub-picosecond temporal resolution[15,16]. Because the excitation wavelength and polarization are defined, PINEM measures the response that governs plasmon-enhanced spectroscopy. PINEM has been applied to the fields at graphene steps[19] and to coupled metal nanoparticles[15,16,20]. It has not been applied to a metal nanoparticle in contact with a 1D vdW crystal.

Here, we report the plasmonic coupling between Au nanoparticles and $NbS_3$ nanowires and nanoribbons, measured by PINEM and applied to the Raman detection of R6G. For PINEM, $NbS_3$ was grown by salt-assisted VLS on graphene-covered TEM grids. Au particles in contact with $NbS_3$ and isolated Au particles on the same support were therefore imaged under identical excitation conditions. Atomic-resolution TEM shows an atomically sharp Au/$NbS_3$ interface. PINEM at 1030 nm shows that an Au particle in contact with $NbS_3$ produces a near field of higher intensity and larger spatial extent than an isolated Au particle with similar radius, and that the near field at a given contact is excited only when the incident polarization is parallel to the particle-$NbS_3$ axis. Finite-difference time-domain (FDTD) simulations reproduce these observations and quantify the dependence of the coupling on the particle-$NbS_3$ separation, the dimension ratio of the $NbS_3$ structure, and its thickness. Finally, colloidal Au nanoparticles (10, 20, and 40 nm) deposited on catalyst-free $NbS_3$ nanowires increase the Raman intensity of R6G at 1362 $cm^{-1}$ by a factor of 3.2 ± 1.4 relative to Au nanoparticles alone, with a detection limit of $10^{-7}$ M. These results extend plasmonic metal/vdW hybrids from 2D layers to a 1D vdW material and relate the measured near-field distribution at the Au-$NbS_3$ contact to the Raman enhancement.

## 2. Results and Discussion

### 2.1 Salt-assisted VLS growth of $NbS_3$ nanowires and nanoribbons

Figure 1a shows the crystal structure of $NbS_3$, in which $NbS_6$ trigonal prisms share faces to form chains along the [010] (**b**) axis. $NbS_3$ nanowires and nanoribbons were synthesized by salt-assisted VLS growth in a horizontal tube furnace (Fig. 1b), following our previous procedure[7]. The synthesis parameters are given in the Methods section. Au films of 1-25 nm thickness were deposited by electron-beam evaporation on $SiO_2$/Si and c-sapphire substrates (Fig. 1b, top) and on few-layer graphene transferred onto holey SiN TEM grids (Fig. 1b, bottom). On heating, the films dewet into particles that serve as the VLS catalysts. A mixture of Nb powder and NaCl was used as the metal source, with the substrate placed face down above it. S was introduced from upstream after the substrate reached the growth temperature of 725-775 °C, with Ar (100-250 sccm) as the carrier gas. Growth on TEM grids allows the Au-$NbS_3$ contacts to be characterized by TEM, STEM, and PINEM in the as-grown state, without transfer.

Figures 1c and 1d show SEM images of growths on $SiO_2$/Si with initial Au film thicknesses of 5 and 15 nm, respectively (see also Fig. SI1). For Au films thinner than about 5 nm, the dewetted particles are too small to sustain VLS growth. $NbS_3$ then grows by the VS mechanism, and the Au particles decorate its surface (Fig. 1c). For Au films of 10–25 nm, VLS growth produces nanowires with a square cross-section that grow along the chain direction [010], and nanoribbons with a rectangular cross-section that grow along [100] (Fig. 1d), as reported previously[7]. The catalyst particle remains at one end of each structure. The initial Au thickness determines the $NbS_3$ morphology, the size of the Au particles, and whether the particles are located at the tip (VLS) or on the surface (VS) of the $NbS_3$ structure. Both configurations of Au-$NbS_3$ contact are examined below.

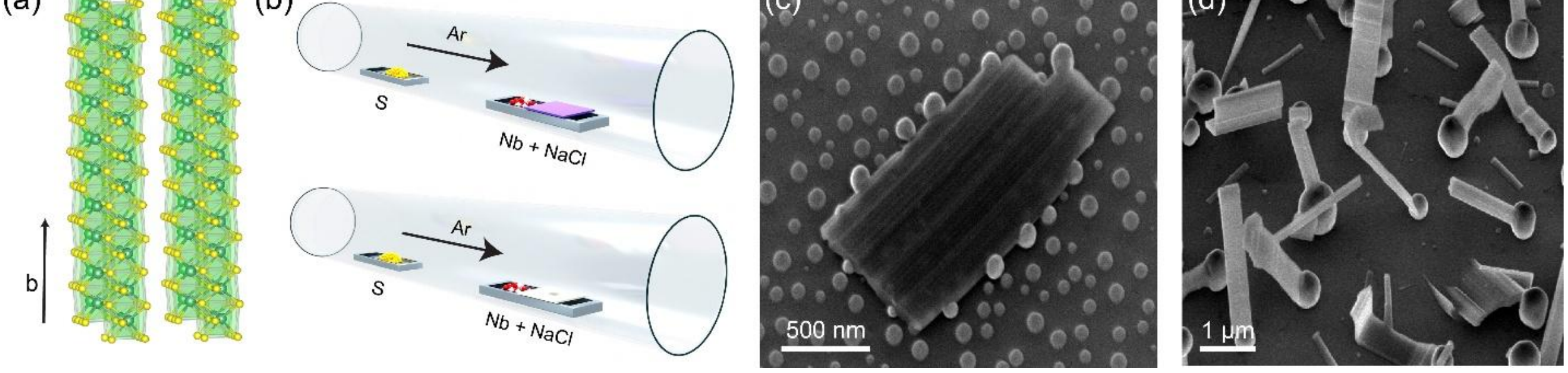

**Figure 1**. Synthesis of Au-$NbS_3$ nanostructures. (a) Crystal structure of $NbS_3$ (Nb, green; S, yellow). The chain direction **b** is indicated. (b) Schematic of the CVD setup for salt-assisted

VLS growth on bulk substrates ($SiO_2$/Si, c-sapphire; top) and on graphene/SiN TEM grids (bottom). (c, d) SEM images of $NbS_3$ grown on $SiO_2$/Si from Au films with initial thicknesses of 5 nm (c) and 15 nm (d).

### 2.2 Structure of the Au-$NbS_3$ junctions

Figure 2a shows a TEM image of $NbS_3$ nanostructures grown on a graphene/SiN TEM grid (see also Fig. SI2). Nanowires and nanoribbons lie on the suspended graphene together with dewetted Au particles with diameters from about 15 to 220 nm. Most 1D structures terminate in an Au particle at one end, which is the signature of VLS growth[7,21], and smaller Au particles are attached to the sidewalls of several structures. Three configurations of Au are therefore present in the same specimen: catalyst particles at the tips of the 1D structures, particles on their sidewalls, and isolated particles on the graphene. STEM-EDS mapping of a nanowire and its catalyst particle (Fig. 2b) shows that Au is confined to the particle, while Nb and S are confined to the nanowire.

Figure 2c shows an atomic-resolution TEM image of the junction between a catalyst particle and a nanowire. The particle is faceted, and the lattice fringes of $NbS_3$ continue to a flat interface with the particle. Figure 2d shows the interface region marked in Fig. 2c together with the FFTs of the particle and of the nanowire. The FFT of the nanowire is indexed to monoclinic $NbS_3$-II along the [201] zone axis, with $d_{010} = 3.3 \pm 0.1$ Å and $d_{-204} = 3.8 \pm 0.1$ Å consistent with the lattice parameters of $NbS_3$-II[7]. The FFT of the particle is indexed to face-centered cubic Au along [111]. The two FFTs give the orientation relationship $NbS_3$ [201] ‖ Au [111] and $NbS_3$ (010) ‖ Au (2-20). An inverse FFT image reconstructed from the Au and $NbS_3$ reflections (Fig. 2d) shows an atomically sharp interface without an intermediate layer.

Figures 2e and 2f display HAADF-STEM images of Au particles on the surface of a $NbS_3$ nanoribbon, the configuration obtained with thin Au films (Fig. 1c, see also Figs. SI2b-c). The nanoribbon in Fig. 2e is about 43 nm wide and is in contact with four Au particles, two at its ends and two on its sidewalls, with diameters between about 10 and 40 nm. Figure 2f shows one sidewall particle at higher magnification. The particle has the shape of a truncated sphere with a flat contact face on $NbS_3$, and its {111} lattice planes, with a spacing of 2.4 Å, are parallel to the interface. The image resolves the atomic columns of Au and $NbS_3$ up to the interface, which is atomically sharp, as found for the catalyst junction in Fig. 2d. The truncated-sphere shape and the flat contact face suggest that the particle was liquid, or had high atomic mobility, when it attached to the $NbS_3$ surface, consistent with the Au-NaCl alloy catalyst proposed for salt-assisted VLS growth[7]. In both configurations, at the tip and on the sidewall, the Au particle is in direct contact with $NbS_3$ across a flat crystalline interface. These contacts are the subject of the near-field measurements in Section 2.3.

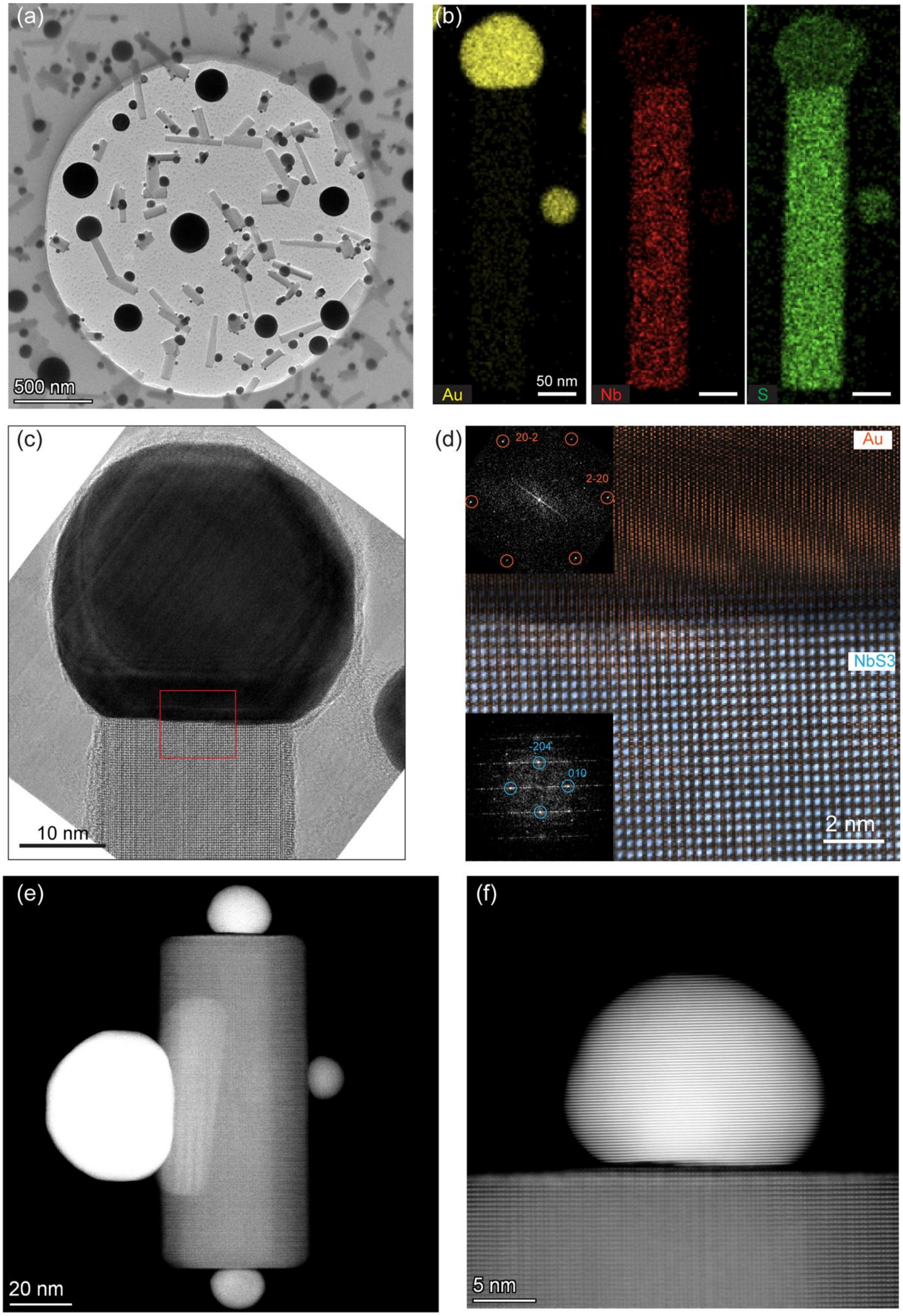
(a)
500 nm
(b)
Au
50 nm
Nb
S
(c)
10 nm
(d)
20-2
2-20
Au
NbS3
-204
010
2 nm
(e)
20 nm
(f)
5 nm

**Figure 2**. Structure of Au-$NbS_3$ contacts. (a) TEM image of $NbS_3$ nanowires and nanoribbons grown on a graphene/SiN TEM grid, with Au catalyst particles at their tips, Au particles on their sidewalls, and isolated Au particles. (b) STEM-EDS maps of Au, Nb, and S for a nanowire with its catalyst particle. (c) Atomic-resolution TEM image of the junction between an Au catalyst particle and a $NbS_3$ nanowire. (d) Inverse FFT image of the region marked in (c), reconstructed from the Au (orange) and $NbS_3$ (blue) reflections. Insets: FFTs of the Au particle (top) and of the $NbS_3$ nanowire (bottom) with indexed reflections. (e) HAADF-STEM image of a $NbS_3$ nanoribbon with Au particles at its ends and on its sidewalls. (f) HAADF-STEM image showing atomically-sharp interface between an Au particle and $NbS_3$.

### 2.3 Real-space imaging of Au-$NbS_3$ near fields

To investigate the electromagnetic coupling in the Au-$NbS_3$ heterostructures, we performed PINEM measurements with the configuration illustrated in Fig. 3a. The optical pump and the electron probe are derived from the same femtosecond (fs) laser, and a motorized delay stage sets the delay between them. Electrons that pass through the optically excited near field exchange energy with it in multiples of the photon energy $\hbar\omega$, so that the electron energy spectrum shows sidebands separated by $\hbar\omega$ on both sides of the zero-loss peak[17,18]. PINEM images are formed with an energy-selecting slit that passes only the electrons that gained energy. Further details are given in the Methods section. The coexistence of Au particles in contact with $NbS_3$ and isolated Au particles on the same support allows both to be recorded in one image under the same pump fluence, electron dose, and alignment.

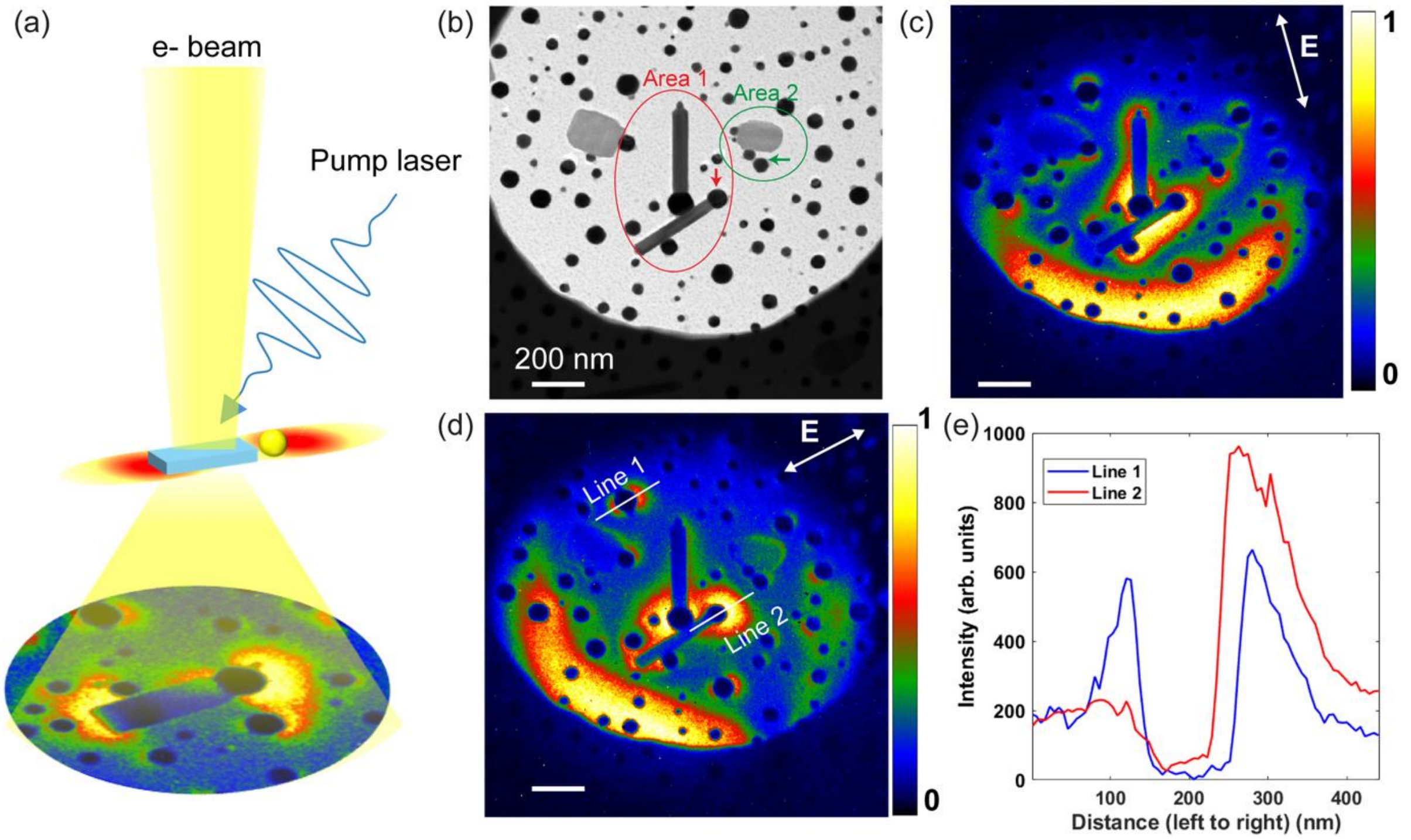

**Figure 3.** Polarization-dependent near-field localization in Au-$NbS_3$ heterostructures. (a) Schematic of the PINEM experiment. (b) Bright-field TEM image of $NbS_3$ nanoribbons and nanowires decorated with Au particles. The two areas analyzed in the text are marked (Area 1, red; Area 2, green), and the arrows indicate the particles discussed. (c,d) PINEM images for linear polarization at −78° (c) and 22° (d), indicated by the white arrows. The crescent-shaped contrast at the periphery arises from the edge of the SiN membrane. (e) PINEM intensity profiles along Line 1 (isolated Au particle) and Line 2 (Au particle in contact with a $NbS_3$ nanowire) in (d). Scale bars: 200 nm. Pump wavelength: 1030 nm; delay: 0 ps.

Figure 3b shows a bright-field TEM image of the analyzed region, which contains $NbS_3$ nanoribbons and nanowires of different width and thickness decorated with Au particles with radii of 20-75 nm. Particles with radii of about 50 and 75 nm are in direct contact with $NbS_3$. Since these radii are much smaller than the excitation wavelength of 1030 nm, the quasi-static approximation gives a first-order estimate of the optical response of the Au particles. In this limit, an isolated Au sphere is described by an induced electric dipole,

$$\mathbf{p} = 4\pi\varepsilon_0\varepsilon_m r^3 \frac{\varepsilon_{\mathrm{Au}}(\omega) - \varepsilon_m}{\varepsilon_{\mathrm{Au}}(\omega) + 2\varepsilon_m}\mathbf{E}_{\mathrm{loc}}$$

where r is the nanoparticle radius, $\varepsilon_m$ is the permittivity of the surrounding medium, $\varepsilon_{Au}(\omega)$ is the frequency-dependent permittivity of gold, and is $E_{loc}$ the local optical field driving the particle[22]. The dipole moment scales with the particle volume, and the dipolar localized surface plasmon resonance (LSPR) occurs near the Fröhlich condition $\mathrm{Re}[\varepsilon_{Au}(\omega)] = -2\varepsilon_m$ [22], which for gold lies in the visible range[23]. Two consequences follow for the present experiment. First, at 1030 nm the Au particles are excited far from their dipolar LSPR. Second, larger Au particles possess larger effective polarizabilities and can therefore generate stronger and more spatially extended near fields.

To examine how the Au-$NbS_3$ heterostructure geometry modifies this picture, we focus on two areas marked in Fig. 3b, Area 1 (red) and Area 2 (green). PINEM images were acquired at zero delay for two polarization angles, -78° and 22° (Figs. 3c and 3d). The strong crescent-shaped contrast at the periphery of both images originates from the edge of the SiN membrane, and all analyses below concern structures far from this edge. The PINEM maps reveal that the near-field localization is not simply confined to individual nanoparticles but emerges as a collective phenomenon, with dipolar hot spots forming across the $NbS_3$-Au heterostructures. In the $NbS_3$-Au heterostructure, the Au nanospheres are strongly coupled to lossy $NbS_3$ nanowires/nanoribbons, so that both the local driving field and the effective plasmonic response are modified by the presence of the anisotropic, absorbing $NbS_3$ host. Furthermore, the larger Au spheres approach a size regime where finite-size effects, including retardation and radiative damping, become non-negligible and can alter both the resonance frequency and the spatial distribution of the near-fields. To account for these effects, we performed full-wave finite-difference time-domain (FDTD) simulations of the imaged geometry, which allow direct comparison with the PINEM maps. Simulations of Area 1 with the Au particles removed and with the $NbS_3$ structures removed are given in Fig. SI3 and separate the contributions of the two components.

A more detailed analysis of individual nanoparticles within the heterostructure provides direct evidence of the anisotropic nature of the coupling. We first consider a nanoparticle located in Area 2, indicated by the green arrow (Fig. 3b), which is in contact with the $NbS_3$ structure from below. The axis defined by this nanoparticle-$NbS_3$ contact is oriented approximately along the −78° direction. Consistent with this alignment, a pronounced PINEM hot spot is observed around this nanoparticle when the incident polarization is set to −78° (Fig. 3c), whereas the same region exhibits negligible enhancement when the polarization is rotated to 22° (Fig. 3d). Conversely, the particle marked by the red arrow in Area 1 (Fig. 3b) is in contact with the tip of a $NbS_3$ nanowire, and its contact axis is nearly parallel to 22° polarization. In this case, a strong, localized hot spot is observed under 22° excitation (Fig. 3d), while no discernible enhancement appears under the orthogonal −78° polarization (Fig. 3c). The excellent agreement between these experimental PINEM observations and the FDTD simulations (Fig. SI3) confirms that the near-field enhancement at a given junction is governed by the relative orientation of the incident electric field with respect to the particle-$NbS_3$ contact axis. In other words, the activation of a specific hot spot is highly polarization-selective, underscoring the pronounced anisotropy of the hybrid coupling.

To quantify the impact of $NbS_3$ coupling on the near-field distribution, we extracted intensity line profiles along the induced dipole direction for two representative Au nanospheres: one that is effectively isolated and one that is coupled to a $NbS_3$ nanowire (Fig. 3d). According to the quasi-static model, the larger isolated sphere (radius 60 nm) should exhibit a stronger near-field response than the smaller coupled sphere (radius 50 nm), due to its larger volume and, hence, larger polarizability. Contrary to this expectation, the PINEM line profiles in Fig. 3e reveal that the coupled nanoparticle exhibits both higher peak intensity (~1.5 times) and a more spatially extended near-field region compared to the larger isolated nanoparticle. This behavior indicates that the $NbS_3$ nanowire does not simply act as a passive dielectric environment but actively participates in shaping and enhancing the local fields. The enhanced localization observed in the coupled configuration is consistent with the formation of $NbS_3$-Au modes.

Comparison of the areas in Fig. 3 and of the additional heterostructures in Fig. SI4 indicates that the dimensions (length, width and thickness) of the $NbS_3$ structure might influence the field localization, but the number of configurations in the present samples is too small to extract a quantitative dependence. We therefore examine these parameters by simulation, varying the particle-$NbS_3$ gap, the dimension ratio (defined as width/length or w/l) of the $NbS_3$ structure, and the $NbS_3$ thickness as illustrated in Fig. 4.

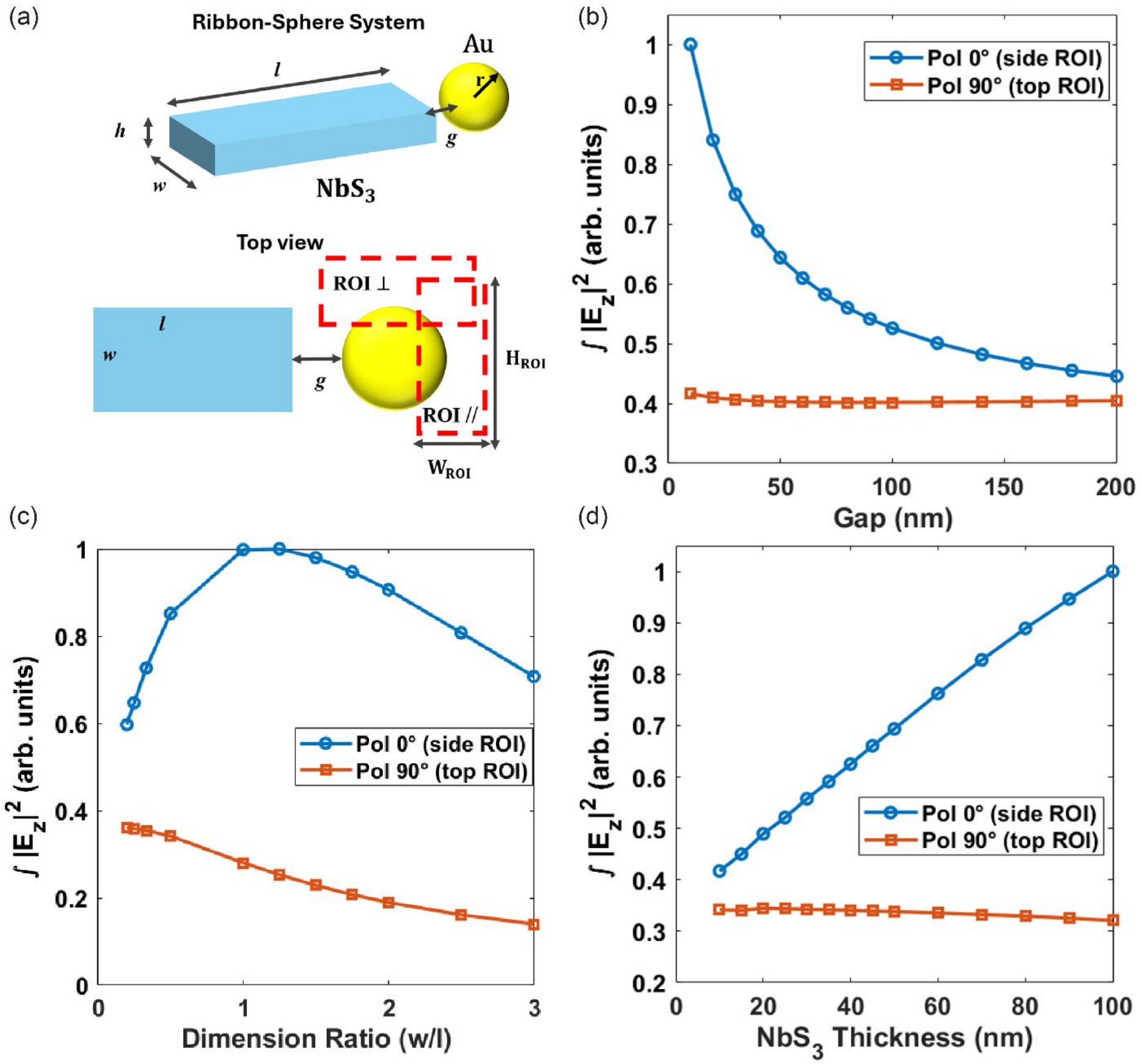

**Figure 4.** Dependence of the Au-$NbS_3$ coupling on the $NbS_3$ geometry (FDTD). (a) Model: a rectangular $NbS_3$ structure of length l, width w, and thickness h at a gap g from an Au sphere of radius 50 nm. Two regions of interest (ROIs; width 40 nm, height 100 nm) around the sphere are used for parallel (0°) and perpendicular (90°) polarization relative to the $NbS_3$-sphere axis, and the integrated $|E_z|^2$ in each ROI is the coupling metric. Nominal values: l = 330 nm, w = 110 nm, h = 50 nm, g = 20 nm and r = 50 nm. (b-d) Integrated $|E_z|^2$ as a function of (b) the gap g, (c) the structure's dimension ratio (w/l) and (d) the $NbS_3$ thickness h. ROI width ($W_{ROI}$) is 40 nm and height ($H_{ROI}$) is 100 nm. Excitation: 1030 nm, normal incidence.

## 2.4 Dependence of the coupling on the $NbS_3$ geometry

We modeled a single $NbS_3$ structure, a rectangular prism of length *l*, width *w*, and thickness *h*, separated by a gap *g* from an Au sphere of radius *r* = 50 nm (Fig. 4a). Unless stated otherwise, l = 330 nm, w = 110 nm, h = 50 nm, and g = 20 nm. The structure is illuminated at normal incidence at 1030 nm, and the polarization angle is defined relative to the $NbS_3$-sphere axis, the line connecting the centers of the $NbS_3$ structure and the sphere. In this configuration, dipolar localized surface plasmons on the Au nanosphere are generated along the incident polarization direction according to the quasi-static description and the PINEM observations in Fig. 3. Therefore, two three-dimensional regions of interest (ROIs) around the sphere, corresponding to parallel (0°) and perpendicular (90°) polarizations, are used to compute the integrated out-of-plane field intensity $\int |E_z|^2 \, dV$ as a measure of coupling strength. g = 20 nm is chosen for the base calculation so that ROI boxes do not overlap with $NbS_3$ region to avoid effects merely arise due to $NbS_3$. Further simulation details are provided in the Methods section.

We first examined the dependence of the coupling on the sphere-$NbS_3$ gap $g$, sweeping from 10 nm to 200 nm (Fig. 4b). Under parallel polarization, the integrated $E_z$ intensity exhibits a strong inverse relationship with the gap. For small separations (10-20 nm), the field intensity is significantly enhanced, indicating strong hybridization between the $NbS_3$ structure and sphere modes. As the gap increases, the coupling progressively weakens and the integrated field converges towards a constant value characteristic of an effectively isolated nanosphere. In contrast, under perpendicular polarization, the integrated $E_z$ shows only a weak dependence on the gap and only for the smallest separations (10-20 nm). Beyond this range, the integrated field remains nearly constant, indicating that the $NbS_3$-sphere pair is essentially uncoupled when the incident polarization is orthogonal to the $NbS_3$ structure-sphere axis.

We next varied the dimension ratio (w/l) of the $NbS_3$ structure (Fig. 4c). For parallel polarization, the integrated field reaches a maximum at an intermediate dimension ratio l/w = 1.25 and decreases on either side. Simulations of the $NbS_3$ structure alone (Fig. SI5) show the same trend, with a maximum of the extinction efficiency at the same dimension ratio. The dependence of the coupled system therefore follows the intrinsic scattering of the $NbS_3$ structure, which is transferred to the sphere by coupling. We attribute this apparent contradiction to the lossy character of $NbS_3$. As seen from the gap-dependence analysis, the ribbon-sphere system is only weakly coupled under perpendicular polarization, so the enhancement contributed by hybridization is small. At the same time, the increasing absorption efficiency and effective area of the $NbS_3$ ribbon with larger dimension ratio provide an efficient loss channel for the localized fields. The ribbon thus acts mainly as an absorber in close proximity to the sphere, draining energy from the near-field region and reducing the integrated field at the ROI despite the increased scattering and absorption of the ribbon itself.

Finally, varying the $NbS_3$ thickness from 10 to 100 nm (Fig. 4d) gives an approximately linear increase of the integrated field with thickness for parallel polarization, consistent with stronger hybridization as the optical cross-section of the $NbS_3$ structure increases. For perpendicular polarization, the integrated field decreases slightly with thickness, again because absorption dominates in the weakly coupled geometry. In summary, the coupling between the Au particle and $NbS_3$ is strongest for small gaps, thick $NbS_3$, and polarization along the particle-$NbS_3$ axis, which is the configuration realized by the catalyst and sidewall particles in Section 2.2.

**2.5 Plasmon-enhanced Raman detection of R6G**

The PINEM measurements and simulations show that the near field of an Au particle is enhanced by contact with $NbS_3$. We tested whether this enhancement increases the Raman scattering of molecules adsorbed at the Au-$NbS_3$ contact, using R6G as the probe molecule (Fig. 5 and Fig. SI6). R6G has well-assigned vibrational bands, and its Raman cross section is resonantly enhanced at 532 nm excitation[11,24]. For this experiment, colloidal Au nanoparticles with diameters of 10, 20, and 40 nm were deposited from solution onto $NbS_3$ nanowires grown on c-sapphire without a catalyst[8]. The Raman intensity of R6G on Au-$NbS_3$ showed no systematic dependence on particle size within the measurement uncertainty (Fig. SI7), and the 40 nm particles were used for the comparison between samples because their size is closest to the particles imaged by PINEM and modeled by FDTD (radii of 20 to 75 nm and 50 nm, respectively) and because 40 nm Au spheres are near their dipolar resonance at the 532 nm excitation wavelength. Colloidal particles were used instead of dewetted Au films because of their narrower size distribution. Four samples were compared: (i) bare c-sapphire, (ii) Au nanoparticles on c-sapphire, (iii) $NbS_3$ nanowires on c-sapphire, and (iv) Au nanoparticles on $NbS_3$ nanowires. R6G was deposited from aqueous solutions with concentrations of $10^{-7}$ to $10^{-4}$ M (about 25 µL per sample), followed by drying at 80 °C. Raman spectra were acquired at 532 nm with 5 mW laser power through a 100× objective. Each spectrum is the sum of 10 acquisitions of 10 s.

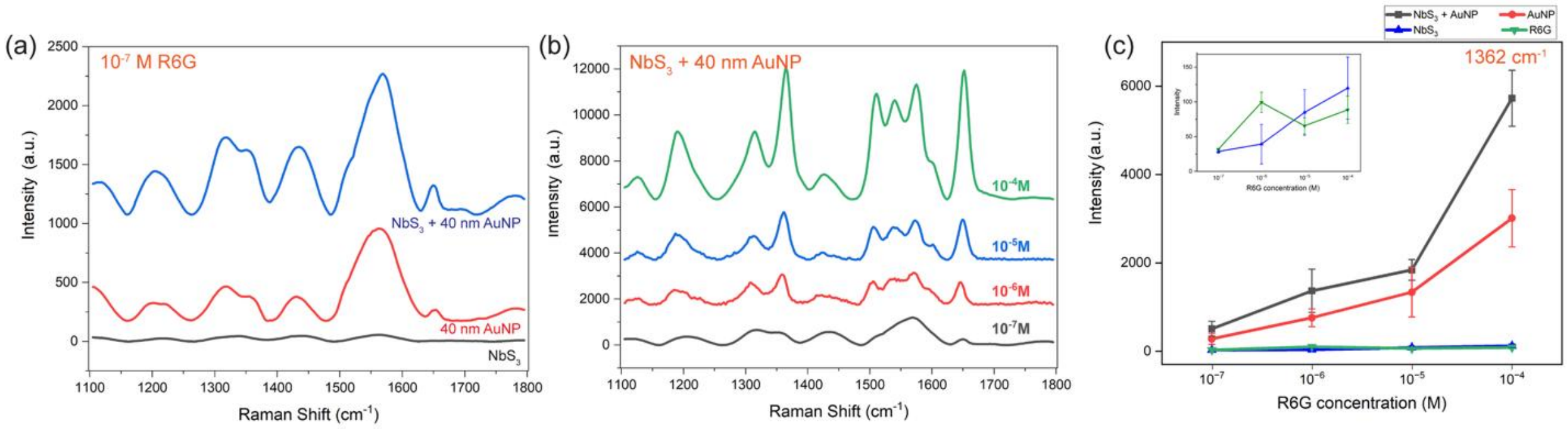

**Figure 5.** Raman detection of R6G on Au–$NbS_3$. (a) Raman spectra of R6G ($10^{-7}$ M) on $NbS_3$ nanowires, on 40 nm Au nanoparticles, and on 40 nm Au nanoparticles deposited on $NbS_3$ nanowires, all on c-sapphire. (b) Raman spectra of R6G on 40 nm Au nanoparticles on $NbS_3$ for concentrations from $10^{-4}$ to $10^{-7}$ M. (c) Intensity of the 1362 $cm^{-1}$ band as a function of R6G concentration for the four samples. Inset: $NbS_3$ and bare c-sapphire on an expanded scale. Error bars are the standard deviation over 3-10 spots.

Figure 5a compares the Raman spectra of R6G at $10^{-7}$ M on $NbS_3$ nanowires, on 40 nm Au nanoparticles, and on 40 nm Au nanoparticles deposited on $NbS_3$ nanowires. R6G bands are present on both Au-containing samples and absent on $NbS_3$ alone, and the intensities are higher on Au-$NbS_3$ than on Au. Figure 5b shows the spectra on Au-$NbS_3$ for R6G concentrations from $10^{-4}$ to $10^{-7}$ M. At $10^{-4}$ M, the characteristic bands of R6G at 1183, 1310, 1362, 1508, 1573, and 1651 $cm^{-1}$ are resolved[24]. As the concentration decreases, the bands broaden and partially merge, but their positions coincide with the resolved bands at $10^{-4}$ M. At $10^{-7}$ M, the peak intensities of the 1310, 1362, and 1573 $cm^{-1}$ bands are about 10-20 times the baseline level between the bands, and we take this concentration as the detection limit under these acquisition conditions. Figure 5c plots the intensity of the 1362 $cm^{-1}$ band as a function of R6G concentration for the four samples. On Au-$NbS_3$, the intensity is 3.2 ± 1.4 times that on Au nanoparticles alone over the whole concentration range. On $NbS_3$ alone and on bare c-sapphire, the intensity is at the noise level (Fig. 5c, inset).

Two mechanisms can contribute to the higher signal on Au-$NbS_3$[11]. The electromagnetic mechanism arises from the plasmonic near field of the Au particle, which is enhanced by contact with $NbS_3$ as shown by PINEM (Section 2.3) and FDTD simulations (Section 2.4). Because the SERS enhancement scales approximately as $|E/E_0|^4$, an increase of the near-field intensity $|E|^2$ at the contact by a factor of ~1.5-2 is sufficient to account for the observed Raman enhancement of 3.2 ± 1.4. The chemical mechanism requires charge transfer between R6G and the substrate. The absence of a measurable signal on $NbS_3$ alone (Fig. 5c, inset) indicates that charge transfer to $NbS_3$ does not contribute significantly under these conditions. The enhancement on Au-$NbS_3$ relative to Au is therefore attributed to the electromagnetic near field at the Au-$NbS_3$ contact.

The detection limit of $10^{-7}$ M does not reach the values reported for optimized metal/2D-material SERS substrates, $10^{-11}$ M for Au nanoparticles on $MoWS_2$ [11] and $10^{-18}$ M for 4H-Au@1T'-$WS_2$ nanowires[14], which contain a high density of hot spots within the laser spot. The present samples were not designed for this purpose. The aim of this work is to connect the structure of the Au-$NbS_3$ contact to its optical response across length scales. Atomic-resolution TEM and STEM resolve the interface (Section 2.2), PINEM images and quantifies the near field at the interface (Sections 2.3 and 2.4), and Raman spectroscopy measures the resulting enhancement over an ensemble of contacts (this section). This combination of methods identifies the Au-$NbS_3$ contact as a plasmonic hot spot in a metal/1D vdW hybrid and provides the basis for substrates with a higher density of such contacts.

## 3. Summary

We have studied the plasmonic coupling between Au nanoparticles and $NbS_3$ nanowires and nanoribbons, a 1D vdW material. Salt-assisted VLS growth on graphene-covered TEM grids produced atomically sharp Au–$NbS_3$ contacts. PINEM at 1030 nm shows that an Au particle in contact with $NbS_3$ produces a near field of higher intensity and larger extent than an isolated particle of similar size, and that the hot spot at a contact is excited only when the incident polarization is parallel to the particle-$NbS_3$ axis. FDTD simulations reproduce these observations and show that the coupling increases with decreasing gap and increasing $NbS_3$ thickness, and depends on the dimension ratio of the $NbS_3$ structure. Colloidal Au nanoparticles deposited on catalyst-free $NbS_3$ nanowires enhance the Raman scattering of R6G by a factor of 3.2 ± 1.4 relative to Au nanoparticles alone, with a detection limit of $10^{-7}$ M, whereas $NbS_3$ alone gives no measurable enhancement.

To the best of our knowledge, this work is the first observation of the plasmonic interaction between a noble metal and a 1D vdW material, obtained by combining atomic-resolution TEM and STEM of the interface, PINEM imaging of the near field at the same interface, FDTD simulations, and Raman spectroscopy of an ensemble of contacts. The coupling is anisotropic and is set by the orientation of the particle-$NbS_3$ axis relative to the incident polarization. We believe the same multimodal approach can be applied to other metal/vdW systems in which the near field at the interface has so far been inferred rather than measured. The 0D/1D heterostructure studied here is the simplest case.

Heterostructures of 1D vdW crystals with each other (1D/1D) and with 2D layers (1D/2D), whose contacts can be defined during growth, offer additional control over the direction and position of field confinement, and the intrinsic anisotropy of the 1D crystal may add to the geometric anisotropy observed here. We anticipate that such dimensionally hybrid systems with growth-defined, atomically sharp interfaces are of interest for polarization-sensitive photodetection[25], nanoscale heat generation[26] and the coupling of plasmons to the collective electronic states of 1D vdW crystals[27].

**Methods**

***VLS Au-NbS$_3$ synthesis:*** SiO$_2$/Si and c-sapphire substrates were cleaned with acetone, isopropanol, and deionized water, followed by oxygen plasma (150 mTorr, 3-5 min). For TEM-compatible growth, few-layer graphene was mechanically exfoliated and transferred onto holey SiN TEM grids. Au films with thicknesses of 1-25 nm were deposited on the substrates by electron-beam evaporation. Au-NbS was grown in 30 minutes by salt-assisted VLS in a horizontal tube furnace as illustrated in Fig. 1b with 50 mg Nb, 25 mg NaCl, 100-150 mg S at 725-775°C, 350-760 torr and 100-250 sccm Ar carrier gas flow. All chemicals were purchased from Sigma-Aldrich.

***VS NbS$_3$ synthesis*:** NbS$_3$ nanowires for the Raman experiments were grown on c-sapphire without Au by salt-assisted chemical vapor deposition. The synthesis parameters are the same as VLS Au-NbS$_3$ synthesis listed above except the Au thin film deposition step.

***Materials Characterization***: SEM images were acquired at 5-15 kV on an FEI Helios Nanolab 600 and a Helios 5 UC. TEM and STEM were performed on a Thermo Fisher Scientific Themis Z and JEOL NEOARM operated at 200 kV. Aberration-corrected HAADF-STEM images were recorded with a semi-convergence angle of 25 mrad and a probe current of 80 pA. FFT analysis, inverse-FFT filtering (Fig. 2d), and EDS processing were carried out with Gatan Microscopy Suite and Velox software.

**PINEM:** PINEM measurements were performed on the ultrafast electron microscope at the Center for Nanoscale Materials, Argonne National Laboratory, a JEOL JEM-2100 Plus modified for ultrafast operation and operated at 200 kV. The output of a femtosecond laser central wavelength 1030 nm, pulse duration 280 fs, repetition rate 250 kHz was split into two beams. One beam was frequency-quadrupled to 257 nm wavelength and directed to the photocathode to generate the probe electron pulses. The other beam (1030 nm) was directed onto the specimen as the pump. Its linear polarization was set by a half-wave plate placed after a linear polarizer. The delay between pump and probe was controlled by a motorized delay stage, and time zero was determined from time-resolved PINEM spectra. PINEM images were recorded at zero delay with a Gatan Imaging Filter, using an energy-selecting slit (10 eV slit width) placed on the energy-gain side of the zero-loss peak. The Au-NbS3 heterostructures were imaged on the graphene/SiN grids on which they were grown, without further preparation.

***Electromagnetic Simulations:*** Electromagnetic simulations of far-field spectra and the near-field distributions were performed using commercial software *Ansys Lumerical FDTD*. The optical constants of NbS$_3$ and gold (Au) were obtained from Dusabe *et al.*[28] and Johnson and Christy's dataset[23], respectively. The geometries and dimensions are extracted from TEM images. Thickness values are obtained through Energy Filtered TEM (EFTEM) thickness mapping. A total-field-scattered-field (TSFS) source was positioned above the structures on the +z side. Perfectly matched layer (PML) boundary conditions were applied along all directions (±z, ±x, ±y). A 3D near-field field and power monitor were placed to encompass the structures and capture the fields. PINEM fields are calculated as $F(x,y) = \int E_z(x,y,z)\, e^{-i(\omega/v)z} dz$ though numerical integration of simulated $E_z$ field, where $v$, the speed of the electron, is assumed 0.7c for our system[18,20]. Additional box monitors are positioned around the structure to record the extinction efficiency.

**Raman Spectroscopy:** Colloidal Au nanoparticles with diameters of 10, 20, and 40 nm suspended in a citrate buffer solution (150 µL) purchased from Sigma Aldrich were drop-cast onto bare c-sapphire and onto c-sapphire with VS-grown $NbS_3$ nanowires. R6G (Sigma Albrich) was dissolved in deionized water at concentrations of $10^{-7}$ to $10^{-4}$ M, and about 25 µL of solution was drop-cast onto each sample and dried at 80 °C on a hot plate. Raman spectra were acquired on a confocal XploRA PLUS Raman microscope with 532 nm excitation, a laser power of 5 mW, and a 100X objective. Each spectrum is the sum of 10 acquisitions of 10 s. Band intensities were determined after baseline subtraction, and the values reported in Fig. 5c are the mean and standard deviation over 5-20 spots per sample.

**Acknowledgements:** The authors acknowledge Frances Ross (MIT) for initiating the collaboration. AN acknowledges the support from ICTAS's Graduate Student Fellowship. Work performed at the Center for Nanoscale Materials, a U.S. Department of Energy Office of Science User Facility, was supported by the U.S. DOE, Office of Basic Energy Sciences, under Contract No. DE-AC02-06CH11357. This work used shared facilities at the Nanoscale Characterization and Fabrication Laboratory, which is funded and managed by Virginia Tech's Institute for Critical Technology and Applied Science. Additional support is provided by the Virginia Tech National Center for Earth and Environmental Nanotechnology Infrastructure (NanoEarth), a member of the National Nanotechnology Coordinated Infrastructure (NNCI), supported by NSF (ECCS 1542100 and ECCS 2025151).